\documentclass[letterpaper,aps,prl,twocolumn,superscriptaddress,showpacs]{revtex4}

\usepackage{graphicx}
\usepackage{amsmath}

\begin{document}

\title{Efficient Fiber Optic Detection of Trapped Ion Fluorescence}

\author{A. P. VanDevender}
\email{aaron.vandevender@nist.gov}
\author{Y. Colombe}
\author{J. Amini}
\altaffiliation{Current Address: Georgia Tech Research Institute, Atlanta, GA}
\author{D. Leibfried}
\author{D. J. Wineland}
\affiliation{National Institute of Standards and Technology, Division 847.10, 325 Broadway St, Boulder, CO 80305}
\date{\today}

\begin{abstract}
Integration of fiber optics may play a critical role in the development of quantum information processors based on trapped ions and atoms by enabling scalable collection and delivery of light and coupling trapped ions to optical microcavities. We trap $^{24}$Mg$^+$ ions in a surface-electrode Paul trap that includes an integrated optical fiber for detecting 280-nm fluorescence photons. The collection numerical aperture is 0.37 and total collection efficiency is 2.1 \%. The ion can be positioned between 80 $\mu$m and 100 $\mu$m from the tip of the fiber by use of an adjustable rf-pseudopotential.
\end{abstract}

%%%% 37.10.Ty Ion Traps
%%%% 03.67.Lx Quantum Computation


\maketitle

%%%%% Introduction

Experiments with single atom, molecule, or quantum dot emitters typically use large multi-element optical systems to collect fluorescence photons. As the emission pattern of such emitters subtends a large solid angle, high numerical aperture (NA) lenses (NA$\approx$1) are used for efficient collection \cite{kimble:prl-39-691,diedrich:prl-58-203,orrit:prl-65-2716,michler:nat-406-968,sortais:pra-75-013406,streed:qic-9-203,shu:0953-4075-42-15-154005}. In some situations, for example, quantum information processing with trapped ions \cite{leibfried:rmp-75-281,blatt:nat-453-1008,haeffner:pr-469-155}, fluorescence detection through optical fibers \cite{weiland:mt-12-965,heine:pra-79-021804,muller:apl-95-173101} may scale to large numbers of ions more easily than bulk objectives as the solid angle access near a trap and field of view are limited. Use of optical fibers may also provide advantages for integrating optical cavities into ion traps because the smaller mode volumes yield stronger coupling of the cavity to the ion \cite{kimble:ps-t76-127,PhysRevLett.83.4987,colombe:nat-450-272}. However, placing a dielectric close to an ion is often problematic, as accumulated charges on nearby dielectrics perturb the trapping potential \cite{PhysRevLett.89.103001,mundt:apb-76-117,keller:jmo-54-1607,george:thesis09}.

\begin{figure}[!ht]
a) \includegraphics[width=2.75in]{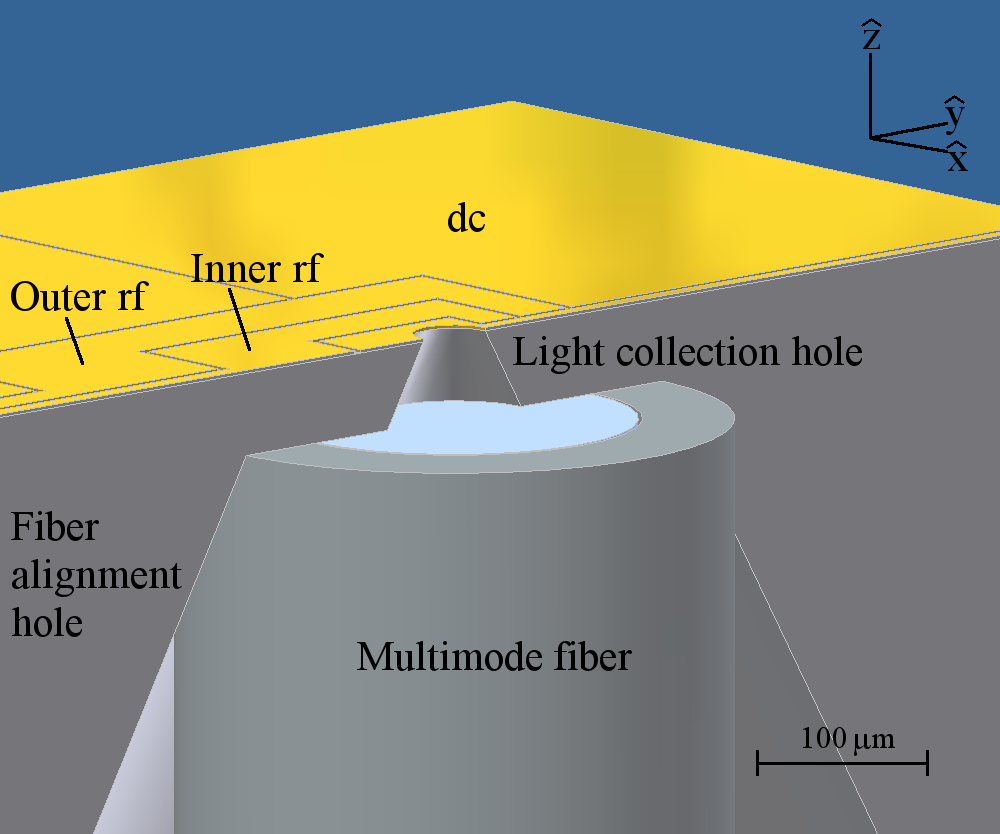}

b) \includegraphics[width=2.75in]{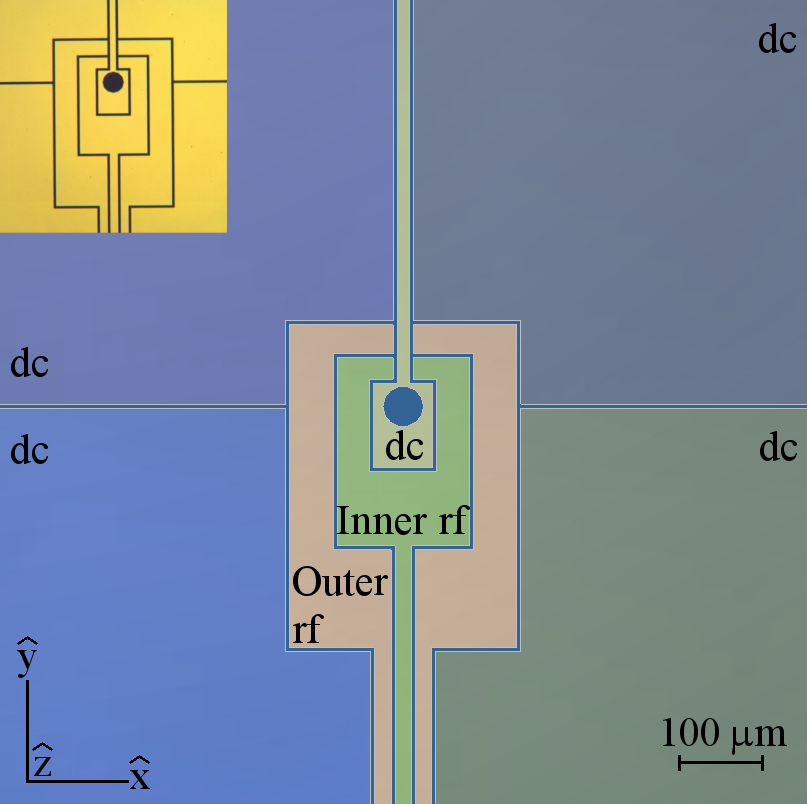}
\caption{a) Trap with electrode pattern cutaway to reveal the tapered light-collection and fiber alignment holes. b) Electrode pattern (photo inset) showing the center dc electrode with light-collection hole, the inner rf electrode, the variable outer rf electrode, and the dc pads.}
\label{FiberInChip}
\label{ElectrodePattern}
\end{figure}

In this letter, we demonstrate a surface-electrode Paul ion trap with an integrated optical fiber used for fluorescence detection. Fluorescence photons pass through a 50-$\mu$m diameter hole in the electrode and substrate layers to be collected in an optical fiber whose face is parallel to and recessed below the electrode surface (Fig.~\ref{FiberInChip}). The other end of the fiber faces a vacuum window with a photomultiplier tube (PMT) on the air side.

%%%%% Construction

The surface-electrode trap is fabricated by use of an electron-beam evaporation and lift-off photolithography process to pattern gold electrodes onto a quartz wafer \cite{Jaeger:2002,amini:njp-12-033031}. The gold layer is 1.5~$\mu$m thick with 5-$\mu$m gaps between the electrodes. A high-speed (150,000 RPM) microdrill that has a diamond-tipped bit with a 30-$\mu$m end diameter and 60$^\circ$ full opening angle is used to mill out a 335-$\mu$m diameter fiber-alignment hole into the back side of the chip. The hole is tapered to allow the fiber to self-align and stop 50~$\mu$m from the electrode surface (Fig.~\ref{FiberInChip}a). A smaller tapered light-collection hole is milled from the fiber recess to within 20 $\mu$m of the front surface. The last 20 $\mu$m of quartz in the 50-$\mu$m diameter light-collection holes is removed using Focused Ion Beam (FIB) milling of the quartz from the front side \cite{langford:mrsb-32} at a 30 degree undercut to match the tapering of the light-collection hole.

The fiber is a multimode high-NA UV-transparent quartz fiber (220-$\mu$m core diameter, 335-$\mu$m cladding outer diameter, 20-cm length) \cite{crystalfiber} with the cladding hole structure in the last $\sim$~40-$\mu$m at each end collapsed to prevent background gasses from slowly escaping from the hole structure. The end surfaces of the fiber are flat. The non-Ultra High Vacuum (UHV) compatible acrylate coatings are removed with dichloromethane, and the fiber is recoated with a polyimide \cite{epo-tek}.

The fiber is inserted into the fiber-alignment hole from the backside and cemented to a bracket on the backside of the chip with an UHV-compatible UV-curing epoxy \cite{optocast}. We verify the alignment of the fiber by sending a 532-nm laser beam into the light-collection hole and checking for light emitted from the opposite end of the fiber.

The trap electrode pattern (Fig. \ref{ElectrodePattern}b) has two independent rf electrodes. An inner electrode provides three-dimensional confinement and a rf pseudopotential null about 30 $\mu$m from the surface plane, centered above the fiber hole, when the outer rf electrode is grounded. When the outer rf electrode is driven at the same potential as the inner, the rf null moves to 50 $\mu$m above the surface plane. This enables vertical control of the ion by scaling the outer rf electrode potential from ground to the same potential as the inner electrode. The range of control could be further extended by grounding the inner electrode and driving only the outer one, which would place an ion 90 $\mu$m from the surface plane. The trap electrodes also include a center dc electrode and four dc pads on the outside of the rf electrodes whose potentials are computer-controlled. These dc electrodes provide a static electric field that cancels any ambient static field at the ion and thereby minimizes rf micromotion \cite{jap:berkeland-83-5025}.

% Loading

Ions are loaded by resistively heating an oven containing a sample of $^{24}$Mg. The oven has been placed 1~cm to the side of the trap and slightly above the plane of the surface electrodes. A very shallow angle is used to minimize plating Mg in the gaps between the electrodes, which could cause shorts. A laser beam at 285~nm photoionizes the atoms, creating $^{24}$Mg$^+$ ions \cite{kjaergaard:apb-71-207}, which are captured in the pseudopotential well.

Ions in the trap are Doppler-cooled \cite{leibfried:rmp-75-281,blatt:nat-453-1008,haeffner:pr-469-155} by use of two laser beams with frequencies tuned near the $^2$S$_{1/2}$-$^2$P$_{3/2}$ transition of the ion (313 nm). The ambient magnetic field is $|\vec{B}| = 0.9$~G. A ``detuned'' beam that has a frequency 300~MHz lower than the atomic transition is used as a first stage of cooling, while a second ``probe'' beam, which has a saturation parameter of 0.2, is tuned $(\Gamma/2\pi)/2{\approx}20$~MHz below the transition for optimal cooling. The Doppler-limited temperature is $\sim$ 1~mK.

Due to the asymmetry of the rf electrodes along the $\hat{y}$ direction, the principal axes of the pseudopotential well are rotated about the $\hat{x}$ axis by 15$^\circ$. This allows for efficient Doppler cooling of all three vibrational modes by use of a single cooling beam aligned in the $(\hat{x}+\hat{y})$ direction. With a 45-MHz drive frequency and a 50-V potential on both of the rf electrodes, the trap motional frequencies are 2.3 MHz, 6.1 MHz, and 8.4 MHz according to our simulation. The lowest-frequency mode is tilted $15^\circ$ up from the $-\hat{y}$ direction, while the highest-frequency mode is tilted $15^\circ$ away from the $\hat{z}$ direction.

%%%%% Resonance

After loading an ion we measure the fluorescence in the following way. We first cool the ion using the detuned and probe beams for 4~ms followed by tuning the probe beam to a detection frequency and applying it to the ion for 400~$\mu$s, during which fluorescence photons are collected. The cooling/detection cycles are performed 4000 times for a given detection frequency to obtain good statistics. By repeating these measurements for a range of detection frequencies around the atomic resonance, we can measure the fluorescence spectrum of the ion.

\begin{figure}[h]
\includegraphics[height=2in]{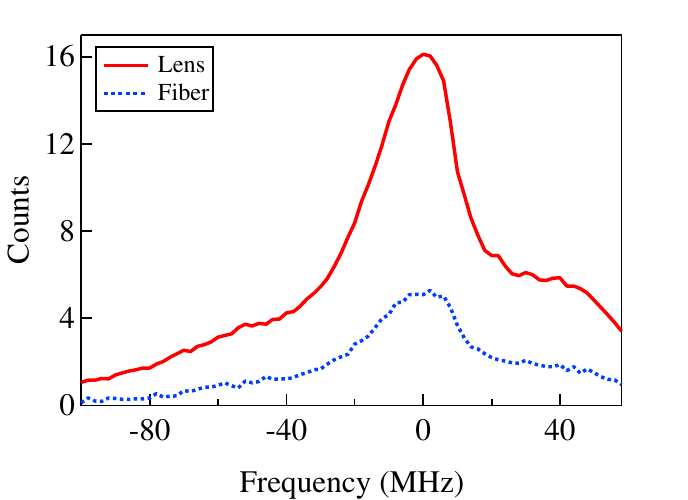}
\caption{Fluorescence counts detected through the lens and the fiber plotted vs. the detuning of the probe laser beam for an ion 50~$\mu$m from the surface plane. The net collection efficiency of the optical fiber is 31 \% of the lens.}
\label{ResonanceScans}
\end{figure}

In addition to the fiber integrated below the trap surface, a large NA=0.5 objective is placed above the trap outside of a vacuum chamber window so that a direct comparison may be made between the fiber collection system and the standard lens collection system that has been used for previous surface ion traps \cite{prl:seidelin-96-253003}.

The fluorescence spectra for the ion when viewed through the lens and the fiber both show the $\sim$~40~MHz linewidth of the resonance (Fig. \ref{ResonanceScans}). The deviation from a Lorentzian profile is due to (1) the frequency dependence of the diffraction efficiency of the acousto-optic modulator used to tune the frequency of the probe light, (2) residual micromotion sidebands \cite{jap:berkeland-83-5025}, and (3) reduced fluorescence due to heating during the detection interval as the probe beam is tuned to frequencies above the resonance. We measure the full resonance spectrum to compare the relative collection efficiencies because the fluorescence scan also reveals the scattered light background of each collection system.

By comparing fluorescence collected with the lens and fiber, we determine the net collection efficiency through the fiber to be approximately 31 \% of the lens efficiency. Although the fiber core is larger than the light-collection hole, some of the light enters the fiber at an angle steeper than the fiber's total internal reflection angle, and is therefore not propagated along the fiber core. Therefore, the net collection efficiency of the fiber is determined by the fiber NA (NA=0.37), reflection losses on the fiber ends and vacuum chamber windows, the 90~\% transmission of a UV filter in front of the PMT, and geometric losses between the fiber and the PMT, because the fiber emits into a cone that is wider than what the active area of the PMT can detect. This geometric loss (coupling coefficient of 0.75) could be eliminated by placing the PMT closer to the fiber; however, the vacuum chamber arrangement in this apparatus prevents that. The lens also suffers from reflection losses on the vacuum window and the objective element surfaces.

%%%%% Position

We verify that the collection efficiency is limited by the NA of the fiber rather than by the size or position of the hole relative to the ion. If the pseudopotential minimum is not well centered over the hole, or if the fiber is not well aligned with the hole, we would expect that displacing the ion laterally would cause the collection efficiency to change as the portion of light clipped by the hole or the fiber cladding changed while the portion limited by the NA would remain constant. By varying the displacement and comparing the dependence of fluorescence through the fiber to the fluorescence through the lens, we may observe any clipping (or lack thereof) in the fiber system and verify the fiber and hole alignment.

\begin{figure}
a) \includegraphics[height=2in]{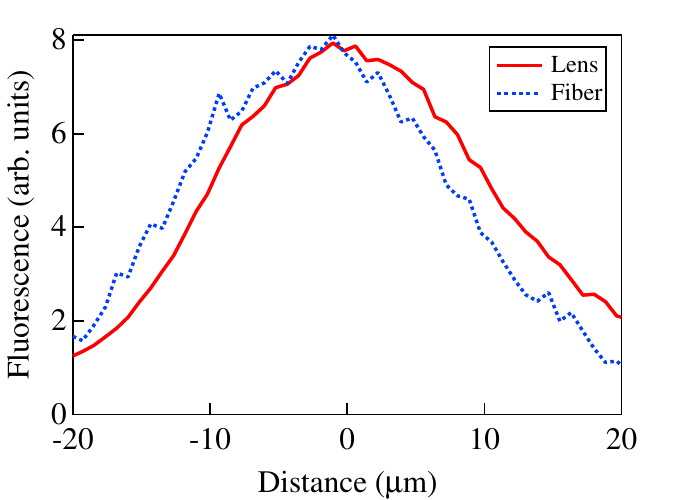}

b) \includegraphics[height=2in]{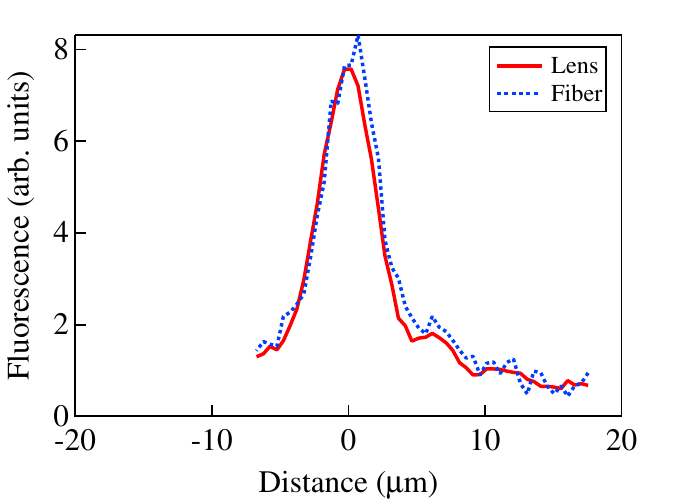}
\caption{Fluorescence observed through the fiber and through the lens when displacing the ion. The fiber trace has been scaled to the height of the lens data. The ion is displaced along the direction of (a) minimum and (b) maximum sensitivity to ion micromotion. The curves are the same for both collection systems, indicating that the fiber collection is limited by its NA rather than clipping by the hole. The displacement between the fiber and lens curves in (a) is due to mechanical instability of the vacuum chamber. For these measurements the probe beam is tuned near resonance.}
\label{Distance}
\end{figure}

Using a numerical simulation of the trapping potential, we find sets of potentials for the dc electrodes that produce displacing fields in three orthogonal directions at the pseudopotential minimum. Because the problem is under-constrained (five electrodes and three dimensions), we use solutions that maximize the displacing field while constraining the electrode voltages to be within $\pm$~10~V. With application of linear combinations of these sets of potentials, the ion may be pushed in any direction. When the ion is pushed in a direction where the rf field at the position of the ion is aligned with the propagation direction of the probe beam, micromotion causes the atomic resonance to be broadened, causing the fluorescence to decrease. Figure \ref{Distance} shows fluorescence as a function of the displacement of the ion in directions that are (a) insensitive and (b) sensitive to micromotion. In the micromotion-insensitive case, the observed changes in fluorescence are due primarily to the ion moving through the intensity profile of the beam. In the micromotion sensitive case, changes are also caused by the induced micromotion. Since the fiber-detected fluorescence is proportional to the fluorescence observed through the lens, we confirm that we are limited only by the NA of the fiber, and not by any misalignment of the light-collection hole or the fiber relative to the ion.

%%%%%% Height

The trap permits adjusting the height of the pseudopotential minimum above the electrode plane by varying the potential of the outer rf electrode. The trap rf is driven with a helical resonator with a $Q$ of $\sim$~40. By tapping into an intermediate position on the helix of the resonator, we can drive the outer rf electrode with a potential between 0~V and that of the inner rf electrode while keeping the phase and frequency of the two electrodes the same. The adjustment could potentially be made continuous with the use of a mechanically adjustable tap or a tunable shunt capacitance divider \cite{cetina:pra-76-041401,herskind:jpb-42-154008}.

We are able to trap ions at 50-$\mu$m above the electrode plane without applying a dc compensation field. A small residual electric field---likely due to photoelectric charging of the quartz substrate and the fiber---displaces the ion from the rf pseudopotential null. Subsequent application of compensation potentials to the dc electrodes cancels the $\sim$~500-V/m ambient field and eliminates excess micromotion in the direction of the probe beam.

At 30 $\mu$m from the surface plane, the combination of decreased trap well depth (3100 K vs. 3800 K), decreased distance of the ion to the stray charges on the surface, and increased stray charges from the detuned and probe beams brought closer to the surface, prevents the ion from loading with all of the dc electrodes grounded. The stray bias field needs to be at least partially canceled out by use of the dc electrodes before the trap will load an ion at this position. We find the collection efficiency of fluorescence photons through the optical fiber at 30~$\mu$m does not improve over the efficiency  at 50~$\mu$m, consistent with the collection being limited by the fiber NA. We are also able to load at $\sim$~40~$\mu$m above the surface by tapping the rf helix at an intermediate point between the grounded end and the high-voltage end, which demonstrates the adjustable property of the technique.

%Conclusion

In conclusion, we have developed a surface-electrode ion trap with integrated fiber optics capable of efficiently collecting fluorescence light from the ion. The trap used here demonstrates the ability to position the ion at a range of distances from the fiber. Although the fiber hole subtends a NA of 0.5-0.8, depending on the ion height, the actual collection is limited to the NA of the fiber (NA=0.37, corresponding to 3.5 \% of the total solid angle). Reflection, absorption, and geometric losses contribute an additional loss of 42 \% of the light, resulting in a net collection efficiency of $\sim$~2.1 \%. By using a similar trap electrode design and shaping the fiber tip to act as a lens to focus the light into the core of the fiber, it should be possible to achieve a much higher effective NA for collection. Furthermore, by anti-reflection coating all interface surfaces and by placing the fiber closer to the PMT to increase the coupling efficiency to the PMT, it should be possible to significantly boost the overall system collection efficiency, making it competitive or superior to conventional collection objectives.

This technique has potential for enabling scalable readout of multiple qubits in trapped ion quantum information processing systems, as fibers may be attached to a single surface-electrode ion trap in multiple locations. This technique is compatible with scalable ion-trap quantum information processing schemes that transport ions with time-dependent control potentials \cite{kielpinski:nat-417-709,hensinger:apl-88-034101,PhysRevLett.102.153002} and may be useful for integrating fibers in neutral atom chip systems \cite{folman:aamop-48-263}. It may also facilitate other quantum optics experiments by, for example, allowing trapped ions to be accurately positioned within a fiber-optic microcavity and enabling strong coupling of the ion to a cavity photon \cite{colombe:nat-450-272}. Dynamically changeable rf-potentials, similar to the ones demonstrated here, could be further leveraged for creating smooth and strongly confining transport potentials through junctions in trap arrays \cite{PhysRevLett.102.153002}.

\begin{acknowledgments}
Supported by DARPA, NSA, ONR, IARPA, Sandia National Labs and the NIST Quantum Information Program. We thank Sandia National Labs for performing the FIB step of our trap fabrication. We also thank John Bollinger and Christian Ospelkaus for their assistance with the experimental apparatus, and Roman Schmied and Ulrich Warring for helpful comments on the manuscript. This letter is a contribution by the National Institute of Standards and Technology and not subject to US copyright. Any mention of commercial products is for information only; it does not imply recommendation or endorsement by NIST.
\end{acknowledgments}

% \bibstyle{unsrt}
% \bibliography{references-short}

\end{document}